\documentclass[11pt]{article}
\usepackage{graphicx}

\title{Monte Carlo Hamiltonian: Inverse Potential\thanks{LXQ 
is supported by the
Key Project of National Science Foundation (10235040), 
National and Guangdong Ministries of Education, 
and Foundation of Zhongshan University Advanced Research Center. HK is supported by by NSERC Canada.}}

\vspace{2cm}
\author{Xiang-Qian Luo$^1$, Xiao-Ni Cheng$^1$, and Helmut Kr\"oger$^2$,  \\ 
$^1${\small\sl Department of Physics, Zhongshan University, 
Guangzhou 510275, China}\\
$^2${\small\sl D\'epartement de Physique, Universit\'e Laval,
Qu\'ebec, Qu\'ebec G1K 7P4, Canada} \\}

\date{\today}

\begin{document}
\maketitle

\begin{abstract}
The Monte Carlo Hamiltonian method developed recently allows to investigate ground state and low-lying excited states of a quantum system, using Monte Carlo algorithm with importance sampling. However, conventional MC algorithm has some difficulties when applying to inverse potentials. We propose to use effective potential and extrapolation method to solve the problem. We present examples from the hydrogen system.
\end{abstract}

\leftline{{\bf PACS numbers:} 02.50.Ng, 03.65.-w, 03.65.Ge}
\leftline{{\bf Key words:} Monte Carlo method, quantum mechanics, computational physics}

\section{Introduction}
\label{sec:Introduction}

There are two standard approaches in quantum theory:
Hamiltonian and Lagrangian formulations.
A comparison is given in Tab.\ref{tab1}.
The Lagrangian formulation is very suitable 
for applying the Monte Carlo (MC) method to
systems with many degrees of freedom,
and in the last two decades, it has been widely applied 
to lattice gauge theory\cite{Creutz,Rothe,Montvay}.
In the standard Lagrangian MC method, however,
it is extremely difficult to compute the spectrum 
and wave function beyond the ground state.

Wave functions in 
conjunction with the energy spectrum contain more physical information 
than the energy spectrum alone. Although lattice QCD simulations 
in the Lagrangian formulation
give good estimates of the hadron masses, one is yet far from a 
comprehensive understanding of hadrons. Let us take as example a 
new type of hardrons made of
gluons, the so-called glueballs. Lattice QCD calculations 
\cite{Luo:1996ha,Luo:96}
predict the mass of the lightest glueball with quantum number 
$J^{PC}=0^{++}$,
to be $1650 \pm 100 MeV$. Experimentally, there are at least two
candidates: $f_0(1500)$ and $f_J(1710)$. The investigation of the 
glueball production and decays can certainly provide additional 
important information for
experimental determination of a glueball. Therefore, it is important 
to be able to compute the glueball wave function.

A natural question is whether in Lagrangian MC simulations 
one can construct an effective Hamiltonian?
We have recently proposed a new approach\cite{mch} 
(named Monte Carlo Hamiltonian method) to investigate this problem.
We start out from the action of the theory.
We compute the transition matrix for all transitions using MC with importance sampling, where the states 
are taken from a basis in Hilbert space. As a result, 
we find an effective Hamiltonian, i.e. its spectrum and eigen states.
A lot of models\cite{mch,physica,lat99_1,Jiang,lat99_2,Huang,Luo:2001fr,Luo:2002rx} 
in quantum mechanics (QM) have been used to test the method.
This method has also been applied to scalar field theories\cite{Luo:2001gb,Huang:1999fn,Kroger:2002pb,Kroger:2002rh}.

Although MC with importance sampling has been very successful for QM with many local potentials, it is very difficult for QM with inverse potentials, due to the singularity. In this paper, we propose effective potential and extrapolation method to solve the problem. The hydrogen system is used as an example.

The rest of the paper is organized as follows. 
In Sect.\ref{sec:Effective Hamiltonian} and Sect. \ref{sec:MatElem},
the basic ideas of MC Hamiltonian are reviewed. In Sect. \ref{sec:error},
we describe a method to analyze the errors of the spectrum and wave functions.
In Sect. \ref{sec:hydrogen}, a method to investigate the inverse potential is presented. The numerical results for the hydrogen system are given in Sect. \ref{sec:results}
and summarized in Sect. \ref{sec:sum}.

\begin{table}
\caption{Comparison of the conventional methods in the standard formulations.}
\begin{center}
\begin{tabular}{|c|c|c|}
\hline
{\bf  Formulation} & {\bf Hamiltonian}  &  {\bf Lagrangian} \\
\hline
Approach & Schr\"odinger Eq.                    & Path Integral\\
         & $H \vert E_n \rangle = E_n \vert E_n \rangle$
                                                &   $\langle O \rangle ={\int [d x] O[x]
                                                    \exp(- S[x]/ \hbar)  \over
                                             \int [d  x] \exp(- S[x]/ \hbar)}$ \\
\hline
Algorithm & Series expansion,               & MC with importance sampling      \\
          & variational,                    &                    \\
          & Runge-Kutta  ...                &                    \\
\hline
Advantage  & Both the ground state,       & It generates the most    \\
           & and the excited states       & important configs.  \\ 
           & can be computed.             & for the measurements.   \\
\hline
Disadvantage & Analytical methods         & It is difficult to study \\
             & are too tedious for        & the excited states.      \\
             & many body systems;         &                          \\
             & Runge-Kutta works          &                          \\
             & only in 1-D.               &                          \\
\hline
            
\end{tabular}
\end{center}
\label{tab1}
\end{table}


\section{Effective Hamiltonian}
\label{sec:Effective Hamiltonian}

Let us discuss the construction of the effective Hamiltonian in 
several steps \cite{mch}. First, consider in quantum mechanics in 1-dim the motion 
of a single particle of mass $m$ under the influence of a local potential 
$V(x)$. Its classical action is given by
\begin{eqnarray}
S = \int dt ~  \left( \frac{m}{2} \dot{x}^{2} - V(x) \right) ~ .
\end{eqnarray}
Given the classical action, one can determine Q.M. transition amplitudes.
Like in Lagrangian lattice field theory, we use imaginary time in what follows. 
We consider the transition amplitude for some finite time $T$ ($t_{in}=0$, $t_{fi}=T$) and for all combinations of positions $x_{i}, x_{j}$.
Here $x_{i}$, $x_{j}$ run over a finite discrete set of points $\{x_{1},\dots,x_{N}\}$ located on the real axis. Suppose these points are equidistantly distributed (spacing $\Delta x$) over some interval $[-L,+L]$. 
The transition amplitudes are given by the (Euclidean) path integral,
\begin{eqnarray}
\label{eq:TransAmpl}
M_{ij}(T)
= \int [dx] \exp[ - S_{E}[x]/\hbar ]\bigg |_{x_{j},0}^{x_{i},T} ,              
\end{eqnarray}
where $S_{E} = \int dt ~  \left( m \dot{x}^{2}/2 + V(x) \right) $ denotes the Euclidean action.
Suppose for the moment that those transition amplitudes were known. 
From that one can construct an approximate, i.e. effective Hamiltonian.
Note that the matrix $M(T)=[M_{ij}(T)]_{N \times N}$ is a real, positive and Hermitian matrix. It can be factorized into a unitary matrix $U$ 
and a real diagonal matrix $D(T)$, 
\begin{eqnarray}
\label{eq:Factor}
M(T)=U^{\dagger}~D(T)~U ~ .
\end{eqnarray}
Because the matrix $M(T)$ can be expressed in terms of the full Hamiltonian $H$
by
\begin{eqnarray}
M_{ij}(T) = \langle x_{i} | \exp (-H T /\hbar) | x_{j} \rangle ~ ,
\end{eqnarray}
the matices $U$ and $D$ have the following meaning,
\begin{eqnarray}
\label{eq:UDInterpret}
&& U^{\dagger}_{ik}=<x_i|E_k^{eff}>
\nonumber \\
&& D_k(T)=\exp (-{E_k^{eff}}T/\hbar) ~ ,
\end{eqnarray}
i.e., the $k-th$ eigenvector $|E_k^{eff}>$ of the effective Hamiltonian $H_{eff}$ can be identified with the $k-th$ column of matrix $U^{\dagger}$ and
the energy eigenvalues $E^{eff}_{k}$ of $H_{eff}$ can be identified with the logarithm of the diagonal matrix elements of $D(T)$.
This yields an effective Hamiltonian, 
\begin{eqnarray}
H_{eff} = \sum_{k =1}^{N} | E^{eff}_{k} > E^{eff}_{k} < E^{eff}_{k} |.
\end{eqnarray}
Note that in the above we have been mathematically a bit sloppy.
The states $| x_{i} \rangle$ are not Hilbert states. We have to replace  
$| x_{i} \rangle$ by some ``localized" Hilbert state. This can be done by introducing box states. We associate to each $x_{i}$ some box state $b_{i}$,
defined by
\begin{eqnarray}
b_{i}(x) =  \left\{ 
\begin{array}{l}
1/\sqrt{\Delta x_{i}} ~~ \mbox{if} ~~ x_{i} < x \leq x_{i+1}
\\
0 ~~ \mbox{else}
\end{array}
\right.
\label{box}
\end{eqnarray}
where $\Delta x_{i} = x_{i+1}-x_{i}$.
When we use an equidistant distribution of $x_{i}$, i.e. 
$\Delta x_{i} = \Delta x$, we refer to the basis of box states as the regular basis.

\section{Matrix elements}
\label{sec:MatElem}

We compute the matrix element $M_{ij}(T)$ directly from 
the action via Monte Carlo with importance sampling. 
This is done by writing each transition matrix element 
as a ratio of two path integrals.
This is done by splitting the action into a non-interacting part and an interacting part,
\begin{eqnarray}
S_E = S_{0} + S_{V} ~ .
\end{eqnarray}
This allows to express the transition amplitude by
\begin{eqnarray}
M_{ij}(T) = M^{(0)}_{ij}(T) ~
\frac{ 
\left.
\int_{x_{i}}^{x_{i+1}} d y 
\int_{x_{j}}^{x_{j+1}} d z
\int [dx] ~ \exp[ - S_{V}[x]/\hbar ] ~ \exp[ -S_{0}[x]/\hbar ] \right|^{y,T}_{z,0} }
{ \left.
\int_{x_{i}}^{x_{i+1}} d y 
\int_{x_{j}}^{x_{j+1}} d z
\int [dx] ~ \exp[ -S_{0}[x]/\hbar ] \right|^{y,T}_{z,0} } ,
\end{eqnarray}
Here $\exp (-S_{0}/\hbar)$ is the weight factor and $\exp (-S_{V}/\hbar)$ is the observable.
$M^{(0)}_{ij}(T)$ stands for the transition amplitude of the noninteracting system, which is (almost) known analytically.
For details see ref.\cite{mch}.
Carrying out these steps allows to construct an effective Hamiltonian, which has turned out to reproduce well low energy physics of a lot of quantum systems
\cite{mch,physica,lat99_1,Jiang,lat99_2,Huang,Luo:2001fr,Luo:2002rx,Luo:2001gb,Huang:1999fn,Kroger:2002pb,Kroger:2002rh}.

\section{Error analysis}
\label{sec:error}

From errors of the matrix elements $\Delta M$ in Monte Carlo simulations, one can obtain the errors of the eigen value $E_n$ and wave function $\psi_n$ of the effective Hamiltonian.
Suppose $H$ is the Hamiltonian of the theory, $H'$ is a perturbation term induced by error in the simulation, then  we have the following equation for the transition matrix element

\begin{eqnarray}
\exp\left(- \frac {(H+H')T}{\hbar}\right)
 \approx  \exp \left(-{HT \over \hbar}\right)
- {T \over \hbar}  \exp \left(-{HT \over \hbar}\right) H^{'} ,
\end{eqnarray}
which implies
\begin{eqnarray}
\Delta M & = & \exp \left( -{(H+H')T \over \hbar}\right) 
- \exp \left(-{HT \over \hbar}\right)
\nonumber \\
& \approx & 
-{T \over \hbar} \exp \left( -{HT \over \hbar} \right) H^{'}
\end{eqnarray}
or
\begin{eqnarray}
H' \approx - {\hbar \over T} M^{-1} \Delta M .
\end{eqnarray}
According to perturbation theory, the first order correction to the eigen value and  wave function is
\begin{eqnarray}
\Delta E_{n} &=&  H^{\prime}_{nn}= -{\hbar  \over T \exp (-E_{n}T / \hbar )} \int dx ~ \psi^{\ast}_{n} \Delta M \psi_{n}
\nonumber \\
\psi'_{n} &=& \sum_{n'\not=n}
  \frac {H^{\prime}_{n'n}}{E_{n}-E_{n'}}\psi_{n'}
\label{e3}
\end{eqnarray} 
where
\begin{eqnarray}
 H^{\prime}_{n'n}=\int dx ~ \psi^{\ast}_{n'}H^{\prime}\psi_{n} .
\end{eqnarray} 
Eq. (\ref{e3}) gives an estimate of the wave function.

\section{Inverse potential: the hydrogen atom}
\label{sec:hydrogen}

The Schr\"odinger equation of a hydrogen is
\begin{eqnarray}
  -\frac {\hbar^{2}}{2m} \nabla^{2}\Psi(r,\theta,\varphi)
+V(r)\Psi(r,\theta,\varphi)=E\Psi(r,\theta,\varphi) ,
\end{eqnarray}
where
\begin{eqnarray}
 V(r)=-\frac {e}{r} .
\end{eqnarray}
Because of the spherical symmetry, the wave function can be factorized as
\begin{eqnarray}
 \Psi(r,\theta,\varphi)=\psi (r)Y(\theta,\varphi) ,
\end{eqnarray}
and the radial wave function satisfies
\begin{eqnarray}
 -\frac {\hbar^2}{2m}\left[r^2 \frac d {dr}
\left(r^2 \frac {d\psi (r)} {dr}\right)\right]
+\left[V(r)+\frac {l \left(l+1\right)\hbar^{2}}{2mr^{2}}\right]\psi (r)=E \psi .
 \end{eqnarray}
Defining a function $u(r)$ by
\begin{eqnarray}
u(r)= r \psi (r) ,
\end{eqnarray}
and considering the case of angular-momentum quantum number zero ($l=0$), one obtains
\begin{eqnarray}
 -\frac {{\hbar}^2}{2m} \frac {d^{2}u}{dr^2}
+V(r)u = Eu
\label{1d}
\end{eqnarray}
which reduces to a one dimensional QM problem. This system has the following analytic result for the energy
\begin{eqnarray}
E_n= - {1 \over 2(n+1)^2},
\label{analytic_energy}
\end{eqnarray}
where $n=0, 1, ...$ is the principal quantum number and $m=1$, $e=1$, $\hbar=1$ have been chosen. The ground state and first excited state radial wave functions are
\begin{eqnarray}
\psi_0 (r) &=& 2\exp(-r) ,
\nonumber \\
\psi_1 (r) &=& {1 \over {\sqrt 2}} \left( 2-r \right) \exp(-r/2) .
\label{analytic_function}
\end{eqnarray}

Our purpose is to investigate the 1-d system NOT by solving Eq. (\ref{1d}),
but by MC Hamiltonian method. Unfortunately, due to the divergence of
$V(r)$ at $r=0$, it is not convenient for MC simulations.

To solve the problem, we replace $V(r)$ by an effective potential with cut-off parameter $R$
\begin{eqnarray}
 V_{R}(r)=\frac  {\exp(2r/R)-1}{\exp(2r/R)+1} V(r) .
\end{eqnarray}
$ V_{R}(r)$ approaches to $V(r)$ when $R \to 0$ and $ V_{R}(r)$ is analytic for any value of finite $R$, which will be used in numerical simulations.

\section{Numerical results}
\label{sec:results}

The coordinates for the box state (see Eq. (\ref{box})) are chosen to be
\begin{eqnarray}
 r_{i}=i\ast dr , ~~~ i=1,2, ...,N.
\end{eqnarray}
The simulation parameters are
$m=1$, $e=1$, $\hbar=1$, $T=0.85$, $dr=0.99$ and $N=100$.
Using the Metropolis algorithm and method described in Sect. \ref{sec:MatElem}, 
we obtain the matrix elements $M_{n'n}$ with potential $V_R(r)$ for some non-zero $R$.
Then we compute the eigenvalues and eigenvectors
using the method described in Sect. \ref{sec:Effective Hamiltonian} and analyze their errors using the method described in Sect. \ref{sec:error}.

\begin{figure}[hb]
\begin{center}
\rotatebox{270}{\includegraphics[width=8cm]{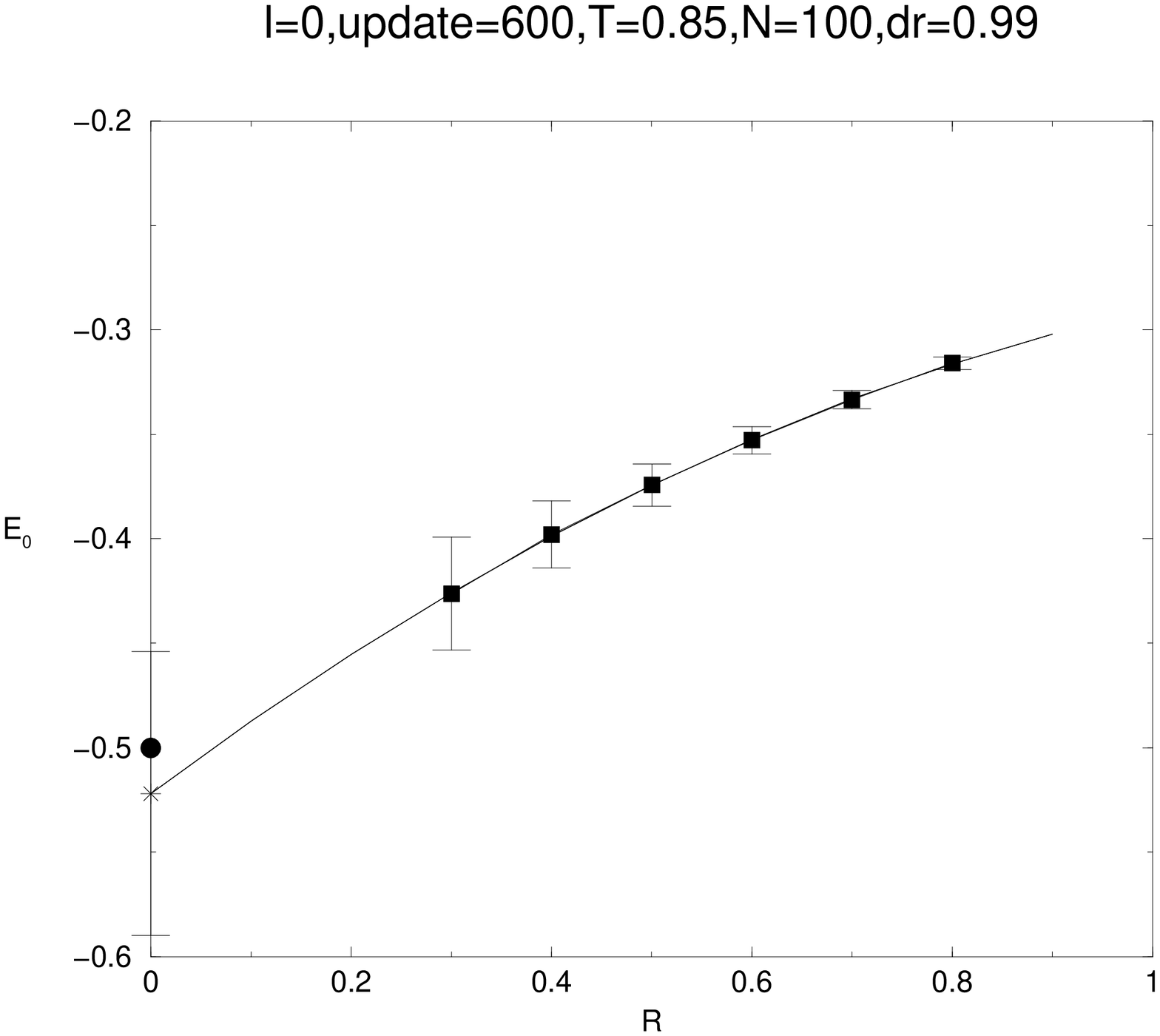}}
\end{center}
\caption{\label{fig1} Ground state energy of the hydrogen system with effective potential $V_R(r)$ at $R=$0.3, 0.4, 0.5, 0.6, 0.7 and 0.8. The circle at $R=0$ is the analytic result and the line is the fitting curve. Between initial and final $r_i$, 600 paths are used in MC simulations.}
\end{figure}

\begin{figure}[hb]
\begin{center}
\rotatebox{270}{\includegraphics[width=8cm]{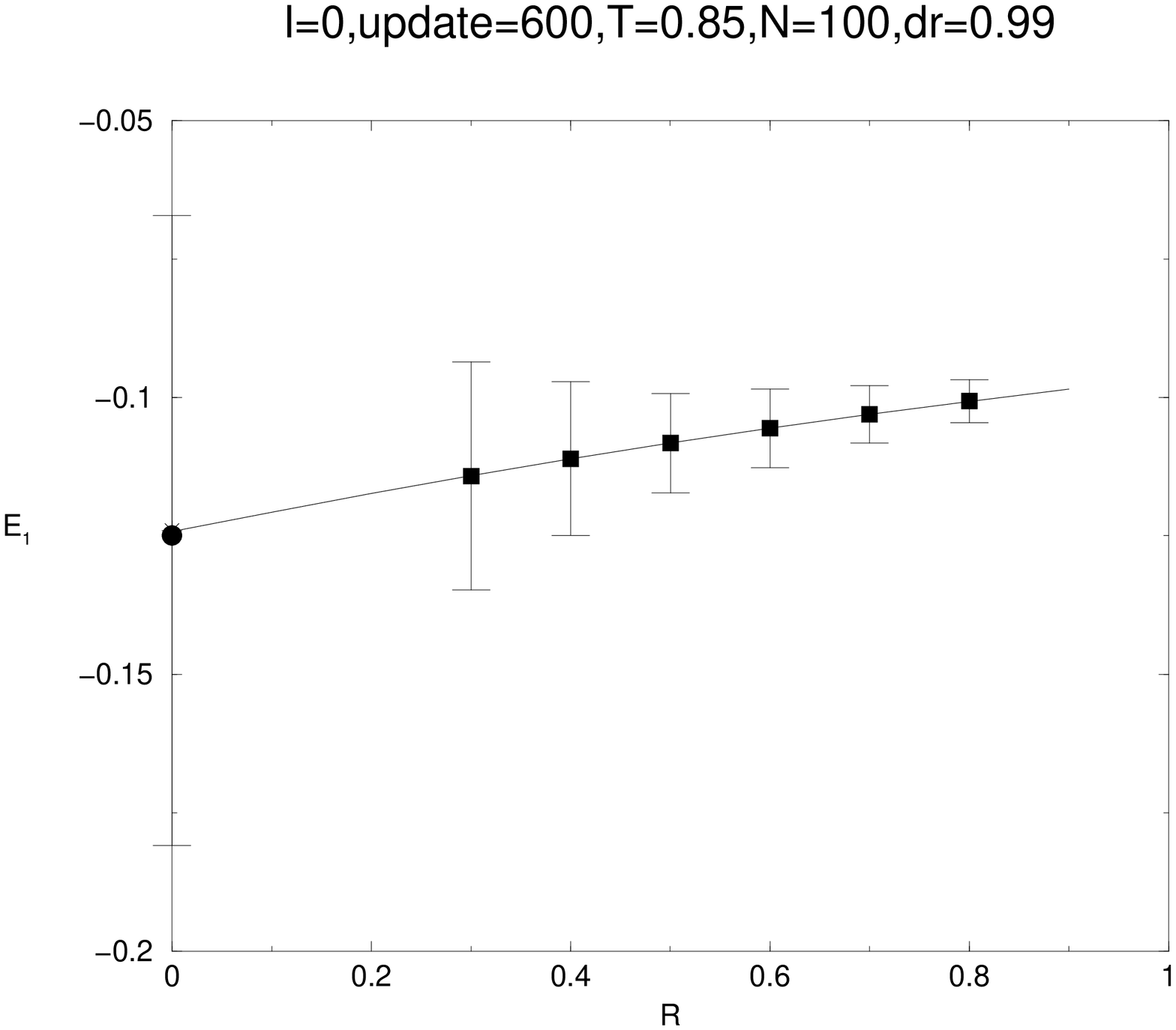}}
\end{center}
\caption{\label{fig2} The same as Fig. \ref{fig1}, but for the first excited state energy.}
\end{figure}

\begin{figure}[hb]
\begin{center}
\rotatebox{270}{\includegraphics[width=8cm]{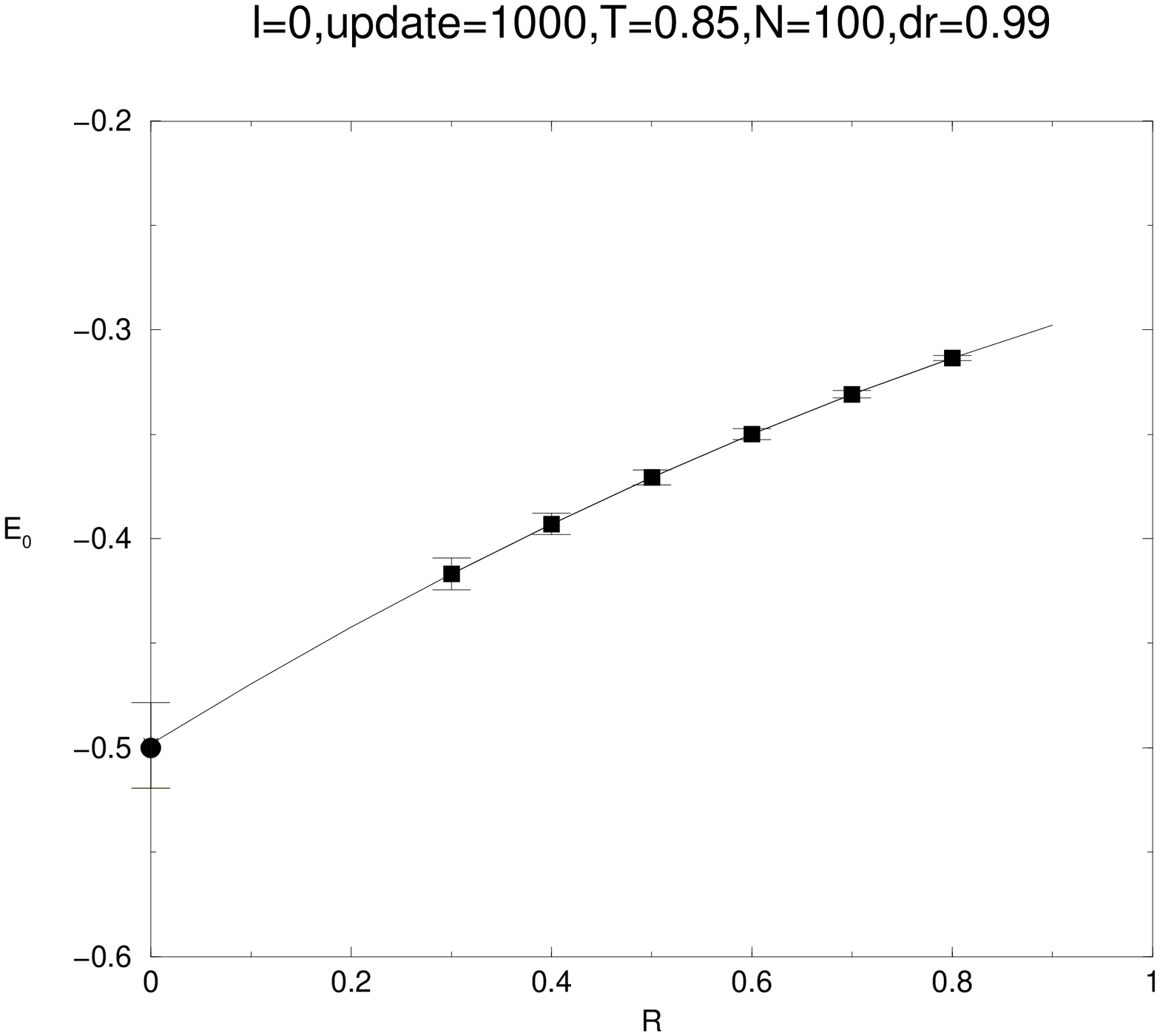}}
\end{center}
\caption{\label{fig3} Ground state energy of the hydrogen system with effective potential $V_R(r)$ at $R=$0.3, 0.4, 0.5, 0.6, 0.7 and 0.8. The circle at $R=0$ is the analytic result and the line is the fitting curve. Between initial and final $r_i$, 1000 paths are used in MC simulations.}
\end{figure}

\begin{figure}[hb]
\begin{center}
\rotatebox{270}{\includegraphics[width=8cm]{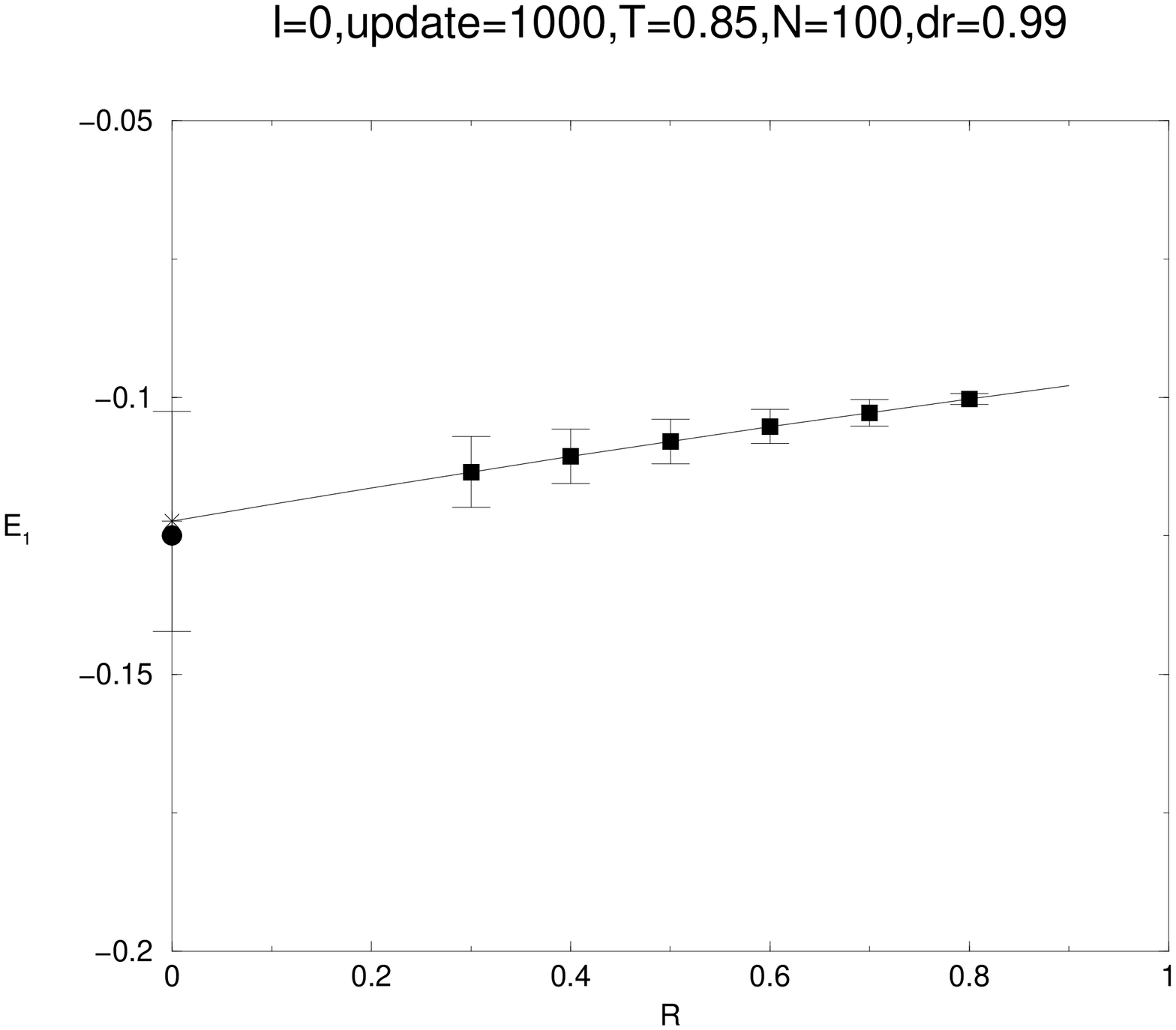}}
\end{center}
\caption{\label{fig4} The same as Fig. \ref{fig3}, but for the first excited state energy.}
\end{figure}

Fig. \ref{fig1} plots the ground state energy of the hydrogen system with effective potential $V_R(r)$ at $R=$0.3, 0.4, 0.5, 0.6, 0.7 and 0.8. 
Between initial and final $r_i$, 600 paths are used in MC simulations.
We use the least square method to
extrapolate the data to the $R=0$ limit. The result agrees with the analytic one 
within large error bar. Fig. \ref{fig2} plots the first excited state energy.
The agreement is greatly improved if more paths (1000) are used in MC simulations. This is shown in Fig. \ref{fig3} and Fig. \ref{fig4}.

\begin{figure}[hb]
\begin{center}
\rotatebox{270}{\includegraphics[width=8cm]{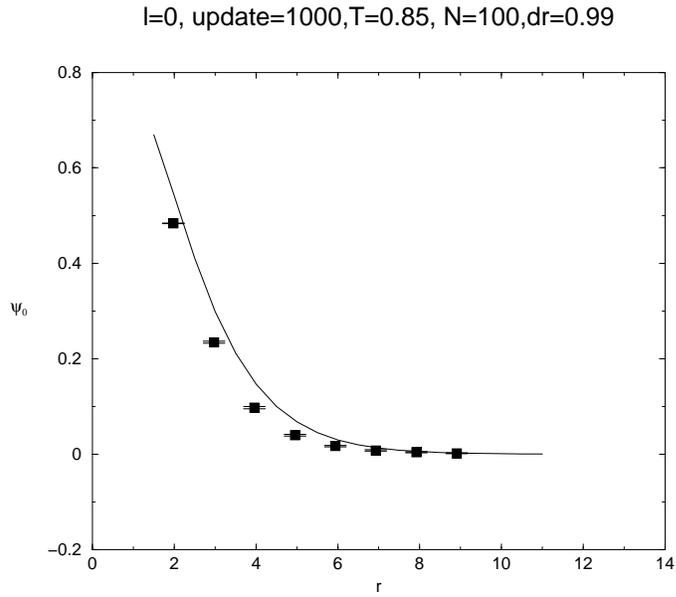}}
\end{center}
\caption{\label{fig5} Ground state wave function of the hydrogen system extrapolated to $R=0$. The line is the analytical result. Between initial and final $r_i$, 1000 paths are used in MC simulations.}
\end{figure}

\begin{figure}[hb]
\begin{center}
\rotatebox{270}{\includegraphics[width=8cm]{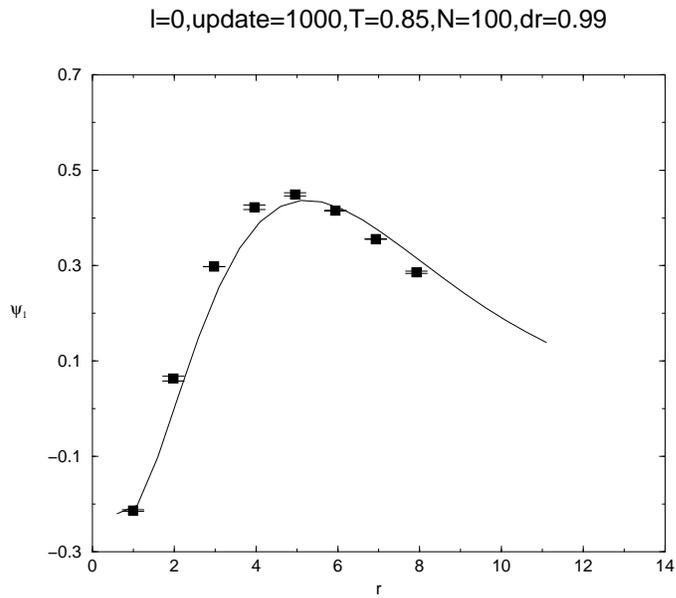}}
\end{center}
\caption{\label{fig6} The same as Fig. \ref{fig5}, but for the first excited state wave function.}
\end{figure}

Fig. \ref{fig5} plots the ground state wave function of the hydrogen system extrapolated to $R=0$. Between initial and final $r_i$, 1000 paths are used in MC simulations. Fig. \ref{fig6} shows the first excited
state wave function. They are consistent with the analytic ones.

\section{Summary}
\label{sec:sum}

In the preceding sections, we have extended the effective Hamiltonian method
to QM with inverse potentials, and taken the hydrogen system as an example.
The results are encouraging. We believe that the application
of the algorithm to more complicated systems will be very interesting. It will allow a non-perturbative investigation of physics beyond the ground state.

\end{document}